# Microdisk modulator-assisted optical nonlinear activation functions for photonic neural networks

Bin Wang,[1,2,3,4] Weizhen Yu, [1,2,3,4] Jinpeng Duan,[3,4] Shuwen Yang,[1,2,3,4] Zhenyu Zhao,[1,2,3,4] Shuang Zheng,[1,2,3,4,*] and Weifeng Zhang[1,2,3,4]

[1]*Radar Research Lab, School of Information and Electronics, Beijing Institute of Technology, Beijing 100081, China*
[2]*Key Laboratory of Electronic and Information Technology in Satellite Navigation (Beijing Institute of Technology), Ministry of Education, Beijing 100081, China*
[3]*Beijing Institute of Technology Chongqing Innovation Center, Chongqing, 401120, China*
[4]*Chongqing Key Laboratory of Novel Civilian Radar, Chongqing 401120, China*
[*]*Corresponding author: zs_bit@bit.edu.cn*

**Abstract:** On-chip implementation of optical nonlinear activation functions (NAFs) is essential for realizing large-scale photonic neural chips. To implement different neural processing and machine learning tasks with optimal performances, different NAFs are explored with the use of different devices. From the perspective of on-chip integration and reconfigurability of photonic neural network (PNN), it is highly preferred that a single compact device can fulfill multiple NAFs. Here, we propose and experimentally demonstrate a compact high-speed microdisk modulator to realize multiple NAFs. The fabricated microdisk modulator has an add-drop configuration in which a lateral PN junction is incorporated for tuning. Based on high-speed nonlinear electrical-optical (E-O) effect, multiple NAFs are realized by electrically controlling free-carrier injection. Thanks to its strong optical confinement of the disk cavity, all-optical thermo-optic (TO) nonlinear effect can also be leveraged to realize other four different NAFs, which is difficult to be realized with the use of electrical-optical effect. With the use of the realized nonlinear activation function, a convolutional neural network (CNN) is studied to perform handwritten digit classification task, and an accuracy as large as 98% is demonstrated, which verifies the effectiveness of the use of the high-speed microdisk modulator to realize the NAFs. Thanks to its compact footprint and strong electrical-optical or all-optical effects, the microdisk modulator features multiple NAFs, which could serve as a flexible nonlinear unit for large-scale PNNs.

Keywords: optical nonlinear activation function, photonic neural network, microdisk modulator.

## 1. Introduction

Over the past few years, artificial neural networks (ANNs) have revolutionized many technical foundations of emerging applications, such as autonomous driving, natural language classification, and medical diagnosis [1-4]. In turn, the computational requirements have escalated rapidly and posed great challenges to the traditional von Neumann computing architecture due to intrinsic bottlenecks in bandwidth and energy efficiency [5-7]. To overcome the limitations, photonic processors have emerged as a promising technology for computing accelerators, capable to provide high bandwidths, high parallelism, low latencies and low crosstalk [8-15]. Particularly, photonic integrated technologies provide a new platform for the implementation of photonic neural networks (PNNs) [16-18].

Recently, various photonic chips have been proposed to implement linear matrix-vector multiplication (MVM) by using microring resonator (MRR) and Mach-Zehnder interferometer (MZI) arrays [8, 10, 12]. Beyond the MVM section, one of the remaining challenges in PNN is the implementation of the nonlinear activation function. In ANNs, nonlinear activation functions (NAFs) are essential, because they allow to model target variables that vary nonlinearly with their explanatory variables, and enable ANNs to learn and perform more complex tasks. Also, different NAFs are required for different areas of neural processing and machine learning tasks with optimal performances. However, in most reported PNNs, the optical NAFs are always executed off-chip due to the absence of integrated optical nonlinear units. To tackle this problem, some integrated approaches have been proposed and demonstrated recently [19-36]. For instance, the photodetector-modulator systems have been demonstrated to exhibit a variety of electrical-optical (E-O) nonlinear transfer functions, which can be employed for different neural-processing tasks [23, 24]. The cavity-loaded MZI structures based on free-carrier dispersion (FCD) and Kerr effect in MRRs have been also used to realize programmable all-optical NAFs [29, 30]. And an MZI mesh-based linear transformer is also proposed to implement multiple types of NAFs [35]. However, the MZI structures can only implement few NAFs, and the large footprints and required power supplies may lead to increased operating costs. In addition, the devices based on germanium and silicon hybrid integration can implement three nonlinear responses [31, 32], while may increase the fabrication complexity. Therefore, from the perspective of on-chip integration and reconfigurability of photonic neural network (PNN), it is highly preferred that a single compact device can fulfill multiple NAFs.

Recently, we reported an add-drop silicon microring to implement all-optical NAFs based on the thermo-optic (TO) effect. However, only few types of NAFs were realized [36]. In this paper, based on our previously reported work, we propose, fabricate and experimentally demonstrate an ultra-compact add-drop microdisk modulator to achieve multiple NAFs. Compared with MZI modulator, microcavity modulator has a smaller footprint and a low optical nonlinear threshold due to resonance-induced energy accumulation. Thus, microcavity modulators offer a great opportunity to realize E-O and all-optical NAFs, and implement more nonlinear responses [23, 32]. In the experiment, a variety of high-speed E-O NAFs can be implemented by electrically controlling free-carrier injection. Moreover, various all-optical NAFs are realized by employing the TO nonlinear effect of the fabricated add-drop microdisk resonator. In the experiment, the nonlinear responses at both through and drop ports could be tuned by controlling the wavelength detuning. The demonstrated nonlinear photonic integration unit may lay the foundation for developing fully on-chip PNNs.

## 2. Operation Principle

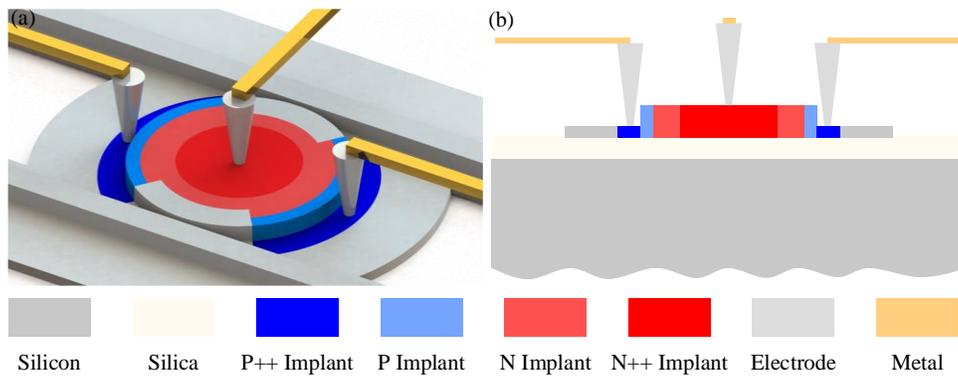

Fig. 1. (a) Schematic of silicon photonic microdisk modulator with a PN junction. (b) A cross-section view of the microdisk modulator.

As shown in Fig. 1(a), the nonlinear activation function is implemented by an add-drop silicon photonic microdisk modulator with a PN junction. The microdisk modulator presented here consists of a microdisk resonator coupled to two neighboring bus waveguides. The cross-section view of the microdisk modulator is shown in Fig. 1(b). In our design, a slab waveguide is introduced to make part of the sidewall further away from the fundamental whispering gallery mode (WGM), which can weaken the scattering loss from the sidewall roughness, and help alleviate the resonance-splitting. Furthermore, the introduction of the slab waveguide also gives a sufficient area to make a PN junction. Reverse-biased PN diodes and forward-biased PIN diodes have been demonstrated to realize high-speed modulation through plasma dispersion effect [37-40]. By tuning the effective index of the waveguide, the resonant shift will induce a strong modulation of the transmitted signal.

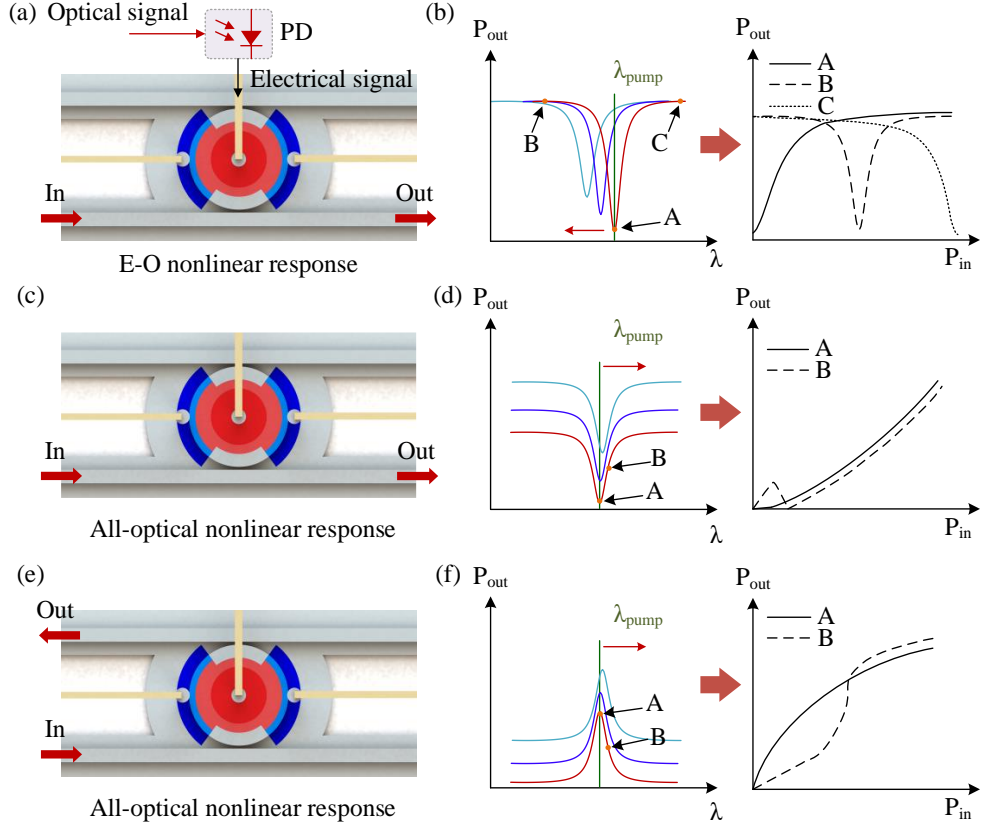

Fig. 2. (a) The schematic of O-E-O nonlinear activation function using a silicon photonic microdisk modulator with a PN junction. (b) The transmission spectra shift and resulted nonlinear responses due to E-O modulation. (c)(e) The schematic of all-optical nonlinear responses at through port and drop port, respectively. (d)(f) The transmission spectra shift and resulted nonlinear responses due to TO effect. The red arrows indicate the direction of the wavelength shift.

The schematic of O-E-O nonlinear activation function using a silicon photonic microdisk modulator with a PN junction is shown in Fig. 2(a). The O-E-O methods convert optical power into an electrical current and then back into the optical signal pathway, and the nonlinear responses usually occur during the E-O conversion process. Thanks to the Lorentzian-shaped transmission spectrum of the microdisk resonator, the nonlinear responses could be derived from the nonlinear E-O transfer function of the modulator. Because reverse-biased PN diodes require a much larger driving voltage than that of forward bias, and the depletion effect is too weak to demonstrate the desired results. Here, we use injection modulation to change the refractive index of guided mode, which can lead to a much larger resonant shift. As shown in Fig. 2(b), with the increase of forward driving voltage, the resonance of the resonator blueshifts due to the accumulation of free carriers in the waveguide and the corresponding decrease of the refractive index of the silicon. In the through output port, different nonlinear responses can be implemented by tuning the driving voltage. For example, when the input wavelength is fixed at the initial resonant wavelength (point A) of the microdisk modulator, the output optical power in the through port will gradually increase as the applied driving voltage increases. When the resonant peak shifts far away from the input optical wavelength, the output optical power will remain unchanged. Thus, a sigmoid-like nonlinear activation function can be realized. Similarly, other relevant nonlinear transfer functions such as radial basis function and negative ReLU can also be implemented by bias tuning [23].

Different from the realization of O-E-O NAFs, all-optical nonlinear responses are dependent on the nonlinear effects of silicon microdisk resonator. The schematic diagram of realizing all-optical NAFs of the through port is shown in Fig. 2(c). The strong light-confinement nature of the silicon microdisk resonator can induce a nonlinear optical response with low input power [41, 42]. In the microdisk resonator, the absorbed optical energy by two-photon absorption (TPA) and free carrier absorption (FCA) is mainly lost by heat, which leads to a redshift of the resonance due to the TO effect. To implement a stable all-optical nonlinear activation function, the wavelength of the input light is fixed at or close to the initial resonant wavelength of the microdisk modulator. In Fig. 2(d), when the input wavelength is set at point A, the output power of the through port first increases linearly with the input power, and there is almost no resonant shift when the input power is relatively weak. When the input power increases, the resonant peak starts to redshift, and the output power increases rapidly due to resonance detuning. As a result, the output power has a Softplus-like nonlinear relationship with the input power. When the input wavelength is set at point B, the output power first increases linearly with low input power. Beyond the threshold, the resonant wavelength starts to redshift and the output power drops due to the notch filtering. With the input power further increases, the output power increases rapidly, finally forming a nonlinear curve RBF in Fig. 2(d).

The schematic diagram of realizing all-optical NAFs of the drop port is shown in Fig. 2(e). In Fig. 2(f), the transmission spectrum of the drop port redshifts due to the TO effect. When the input wavelength is set at point A, the output power first increases linearly with low input power, since there is almost no resonant shift. When the input power continues to increase, the resonant peak starts to redshift, and the output power increases slowly due to resonance detuning and nonlinear loss. As a result, the output power has a saturated nonlinear relationship with the input power in Fig. 2(f). When the input wavelength is set at point B, the output power first increases linearly, and followed by a rapid rise as the input power increases. When the input power continues to increase, the resonant peak deviates from the input wavelength, which makes the output power increases slowly, resulting in a nonlinear relationship in Fig. 2(f). Notably, since the input light wavelength is aligned with or close to the initial resonant peak, the optical bistability could be neglected in the proposed approach and stable nonlinear responses could be implemented [32, 41, 50].

## 3. Device Fabrication and Characterization

Figure. 3(a) shows the microscope image of the fabricated add-drop microdisk modulator used in the experiment. The device is fabricated on a standard 220-nm silicon-on-insulator (SOI) platform. The microdisk resonator has a shallow-etched slab waveguide to surround the disk and the lateral sides of the bus waveguides.

The incorporation of a lateral PN junction is used to achieve the electrical tunability of the microdisk modulator. To optimize the tuning efficiency, the overlap between the PN junction and the fundamental WGM is maximized by slightly shifting the center of the PN junction inward by 500 nm from the edge of the microdisk. Most of the disk is designed to be p-type doped, while the edge of the disk and the slab waveguide are n-type doped since the plasma dispersion effect is more sensitive to the change of the free-hole concentration. In addition, there is no doping along the arc with an angle of 90° near the coupling region, to ensure that the doping does not deteriorate the optical coupling between the bus waveguide and the disk. To satisfy the phase-matching condition for optical coupling between the bus waveguide and the microdisk, the effective refractive index of the fundamental $TE_0$ mode supported by the bus waveguide is required to be equal to that of the fundamental whispering gallery mode (WGM) supported by the microdisk resonator. The bus waveguide is designed to have a width of 600 nm and the microdisk has a radius of 10 μm. The coupling gap has a width of 200 nm. As shown in Fig. 3(a), four $TE_0$-mode grating couplers with a center-to-center space of 127 μm are used to couple light between the chip and the input/output fibers. The grating coupler has a period of 655 nm.

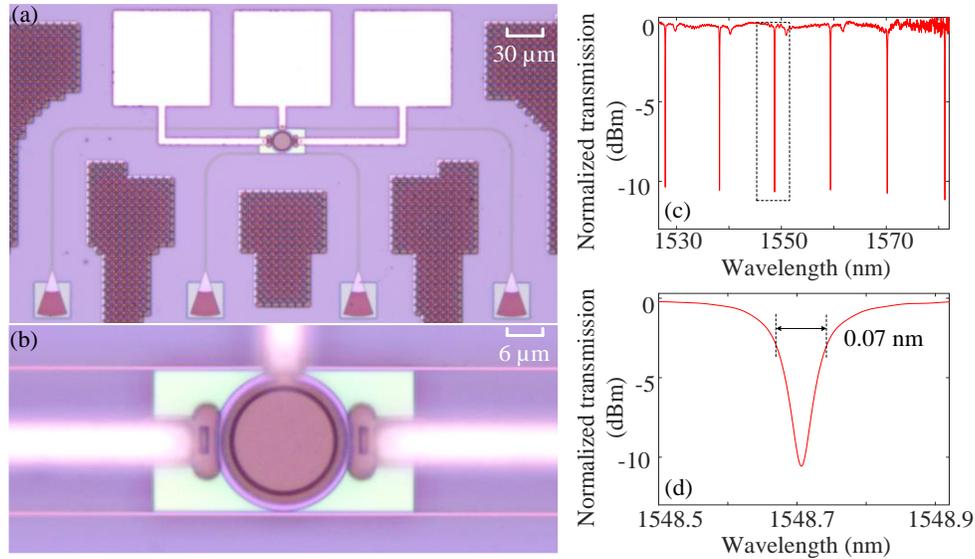

Fig. 3. (a) Microscope image of the silicon microdisk modulator used in the experiment. (b) Details of the fabricated microdisk modulator. (c) The normalized transmission spectrum of the microdisk resonator. (d) Zoom-in view of one single resonance around 1548.71 nm.

Figures. 3(b) shows the zoom-in view of the microdisk resonator. The performance of the fabricated microdisk is first evaluated by using an optical vector analyzer (LUNA OVA). Figure. 3(c) shows the measured transmission spectrum with zero bias voltage. The free spectral range (FSR) is estimated to be 10.5 nm. Figure. 3(d) shows the zoom-in view of one single resonance, and its resonant wavelength is at 1548.71 nm. The resonance has a 3-dB bandwidth of 70 pm, a Q-factor of 22000, and an extinction ratio of 11 dB. The insertion loss of the device is measured to be ~13 dB around 1550 nm, most of which is caused by the fiber-to-fiber I/O coupling loss.

As shown in Fig. 4(a), we further evaluate the performance of the fabricated microdisk modulator when the PN junction is reverse biased. With the increase of reverse-biased voltage, more free carriers are extracted and the depletion region is widened. As a result, the effective refractive index of the waveguide mode is increased, which leads to a redshift of the resonance. In the meanwhile, the decrease in the number of free carriers would reduce the free-carrier induced absorption loss, which could further enhance the Q-factor of the microdisk, and the extinction ratio is also improved since the coupling condition is reaching the critical condition. When the reverse-biased voltage is 8 V, the resonant wavelength is redshifted by 25 pm, and the extinction ratio is increased by 0.8 dB. Figure. 4(c) shows the tuning of the transmission spectra, which gives a wavelength shift rate of 2.4 pm/V. The measurement results for the PN junction being forward biased are shown in Fig. 4(b). When the forward-biased voltage increases, the refractive index of the waveguide mode decreases due to the free-carrier injection. Thus, the excess absorption loss becomes larger, which would degrade the performance of the fabricated microdisk modulator. As can be seen in Fig. 4(b), the resonance blueshifts with the Q-factor and extinction ratio significantly reduced. For example, when the forward-biased voltage is 1 V, the resonant wavelength is blueshifted by 202 pm. The Q-factor is reduced to 18400 and the extinction ratio is decreased by 4.5 dB. Figure. 4(d) shows the tuning of the transmission spectra, which gives a wavelength shift rate of −1.74 nm/V after the PN junction is turned on. Apparently, reverse-biased PN diodes require a much larger driving voltage than that of forward bias.

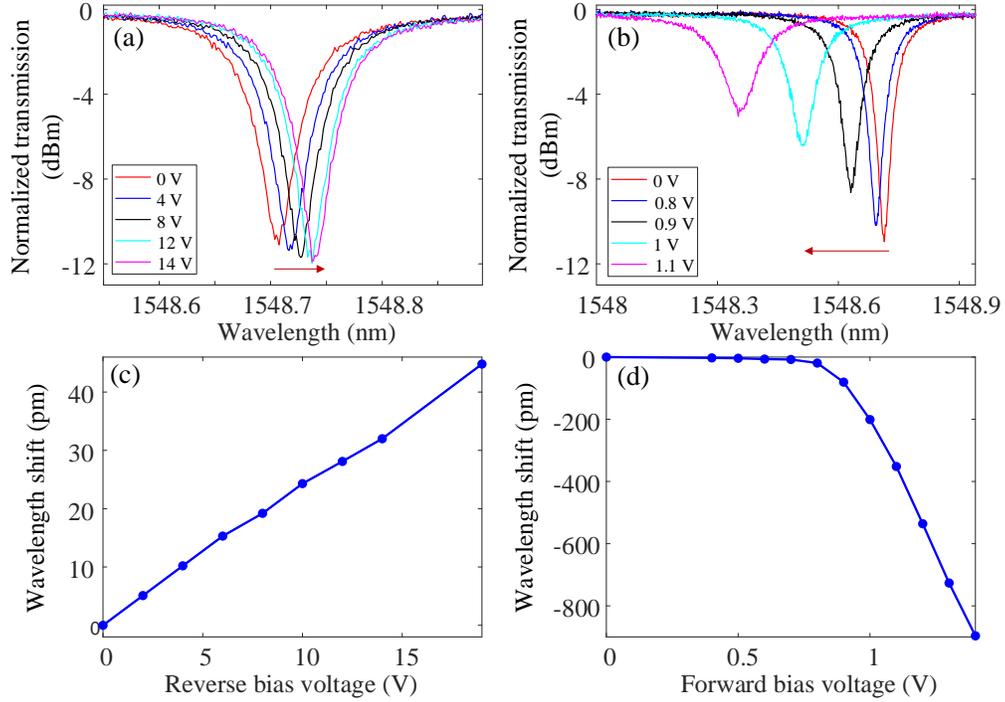

Fig. 4. (a)(b) Measured transmission spectra when the PN junction is reverse and forward biased. The red arrows indicate the direction of the wavelength shift. (c)(d) Measured wavelength shift as the reverse and forward bias voltages change.

## 4. Experiment and discussion

We carry out an experiment to measure the E-O NAFs based on the fabricated microdisk modulator. In the experiment, we use an external voltage source to drive the fabricated microdisk modulator and measure the output temporal signal. The fabricated device is mounted on a micro-positioning stage, and the temperature is kept at approximately 26 °C controlled by a temperature controller (TEC). The RF electric signal from an arbitrary waveform generator (AWG) is then loaded onto the fabricated device through a high-speed microwave probe. The output light is detected by a high-speed photodetector (PD), and then sent into an oscilloscope to monitor the real-time waveforms. To implement various NAFs, the input light is modulated by a triangle waveform at 2.5 MHz, derived from an AWG.

The type of nonlinear response depends on the bias condition and the input light wavelength. As shown in Fig. 5, three typical nonlinear response shapes are realized under different forward bias conditions, each relevant in different areas of neural processing and machine learning [43-45]. In Figs. 4(b) and 4(d), the forward-biased PN junction can be turned on at about 0.7 V. As the driving voltage continues to increase, the resonance blueshifts with both Q-factor and extinction ratio reduced. In Fig. 5(a), the input electrical signal applied on the fabricated modulator has an amplitude of 0.6 V and an offset of 1 V. The input wavelength is located at the resonant peak (~1548.71 nm). With the increase of the applied voltage, the resonant peak gradually moves away from the input light wavelength. Thus, the output optical power in the through port gradually increases and then remains constant. The output electrical signal from PD is recoded by an oscilloscope, which shows a periodic nonlinear curve. The corresponding input and output voltage response curves are presented in Fig. 5(b). Such a sigmoid-like function is commonly used in recurrent Hopfield networks for nonlinear optimization [43].

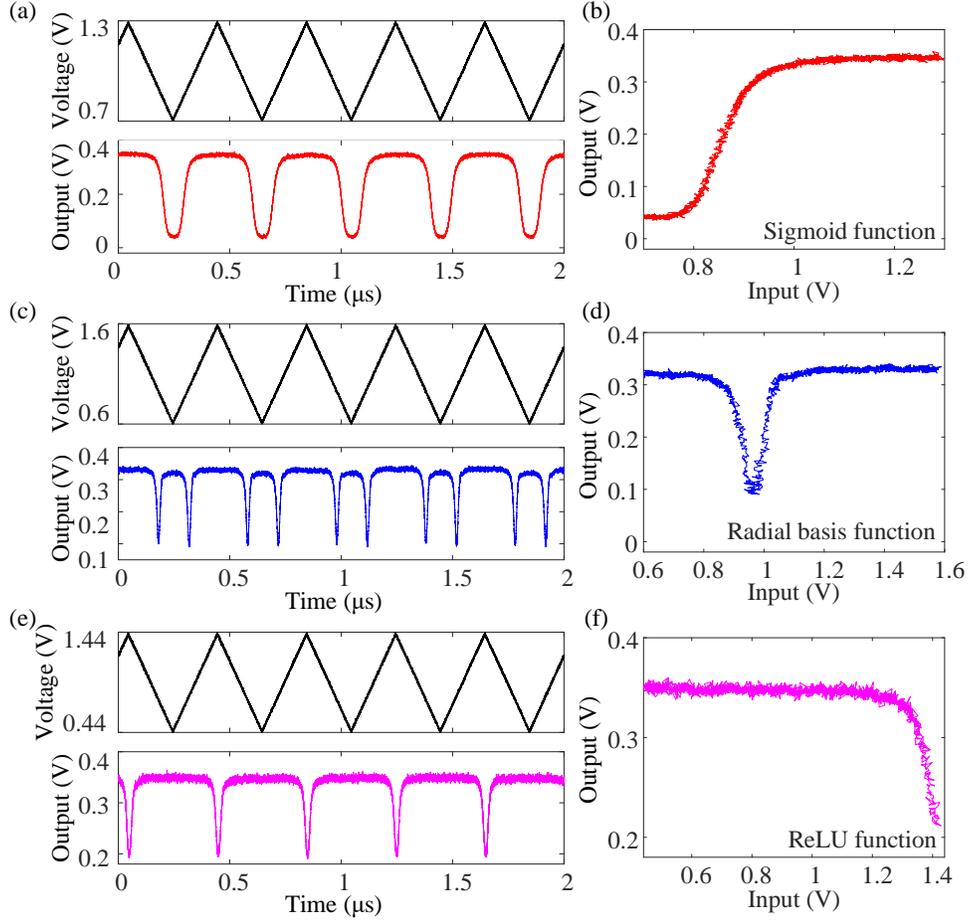

Fig. 5. A variety of relevant nonlinear transfer functions from the modulator, taken at different forward bias conditions: (a)(b) sigmoid function, (c)(d) radial basis function, (e)(f) negative ReLU function. The frequency of the input signal is 2.5 MHz.

In Fig. 5(c), the input electrical signal applied on the fabricated modulator has an amplitude of 1 V and an offset of 1.1 V. The input wavelength is set to be 1548.61 nm, which is blue-detuned. As the driving voltage increases, the output optical power in the through port gradually reduces to the minimum value and then increases due to the blue shift of the resonance. As a result, the radial basis function is obtained in Fig. 5(d), which is commonly used for ML based on support-vector machines [44]. In Fig. 5(e), the electrical signal applied on the fabricated modulator has an amplitude of 1 V and an offset of 0.94 V. The input wavelength is set to be 1547.8 nm, which is far away from the resonance. As the driving voltage increases, the output optical power in the through port remains the same at the beginning and then decreases due to the blue shift of the resonance. As a result, a negative ReLU shape is obtained in Fig. 5(f).

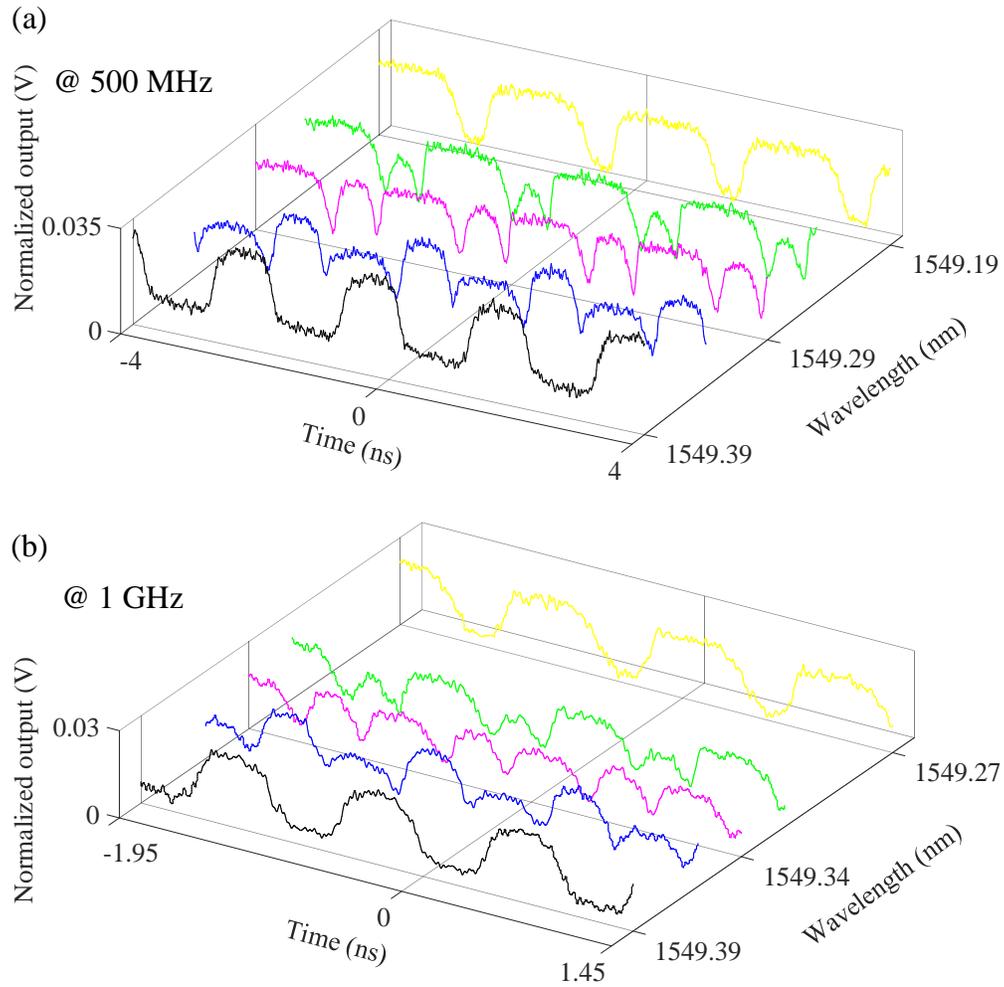

Fig. 6. Measured output waveform at different wavelengths. (a) The frequency of input signal is 500 MHz, and the bias voltage is 0.6 V. (b) The frequency of input signal is 1 GHz, and the bias voltage is 0.6 V.

Then, we measure the high-speed E-O nonlinear responses. Here, the initial resonant peak of the microdisk resonator is around 1549.39 nm. As shown in Fig. 6(a), the frequency of the input triangle waveform is set to be 500 MHz. The bias voltage is 0.6 V and the input electrical signal applied on the fabricated modulator has an amplitude of 1 V. With the change of input wavelength, different waveforms are measured. For example, when the input wavelength is set to be 1549.39 nm, the black curve in Fig. 6(a) is obtained, which shows a periodic sigmoid-like nonlinear curve. When the input optical wavelength changes to 1549.29 nm and 1549.19 nm, the radial basis and ReLU functions are implemented. Similar results at the frequency of 1 GHz are shown in Fig. 6(b). It is apparent that the different nonlinear responses correspond to different pieces of the microdisk resonator's Lorentzian shape, and more response shapes can be implemented by tuning the bias voltage and wavelength. The implemented nonlinear E-O transfer functions are relevant for a wide variety of neural-processing tasks. Since photodetectors' O-E response is linear, a high-speed photodetector can be further integrated with the modulator to realize chip-scale reconfigurable O-E-O NAFs [46-49].

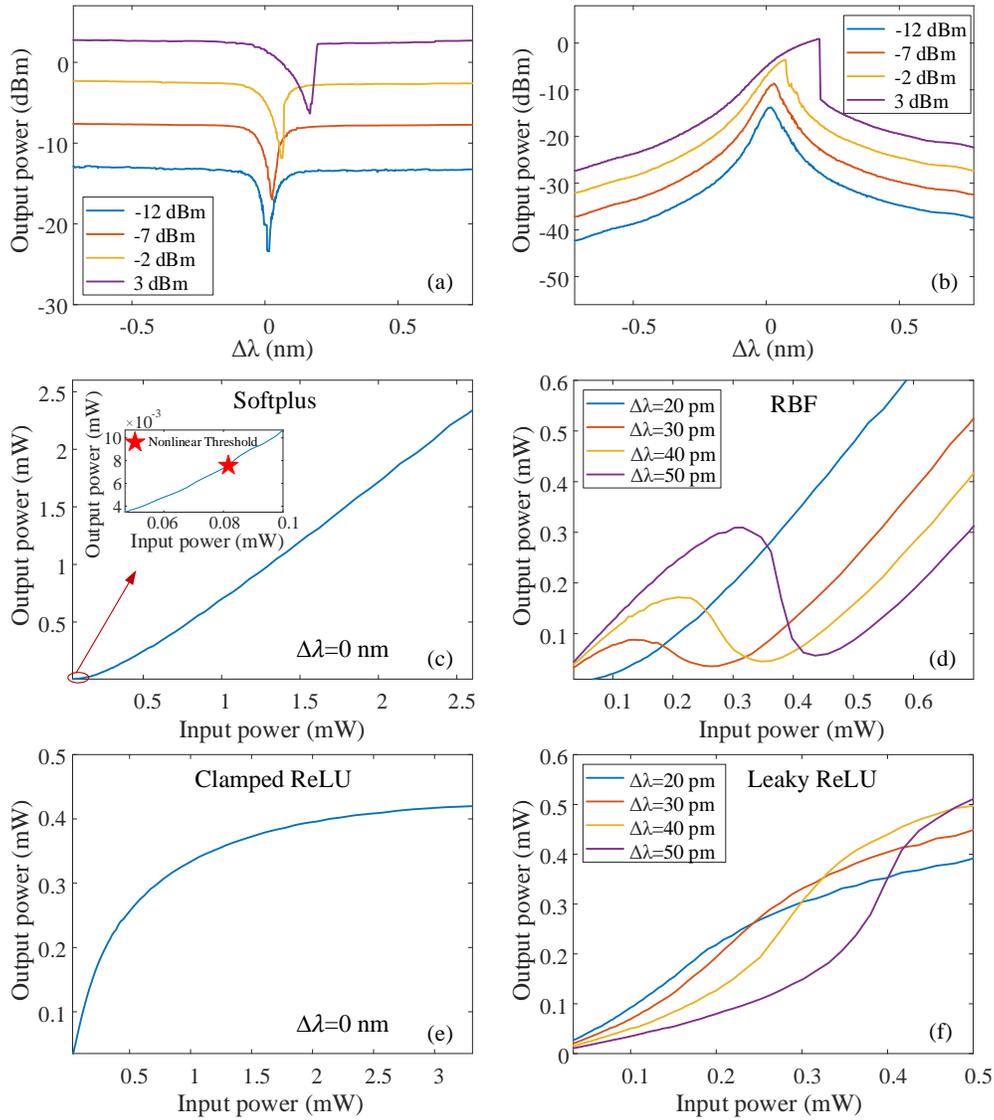

Fig. 7. (a)(b) Measured transmission spectra of the through and drop ports with different input power. (c)(d) Measured nonlinear transfer functions of the through port at different wavelengths. (e)(f) Measured nonlinear transfer functions of the drop port at different wavelengths.

In addition, the fabricated microdisk modulator can be used to realize different all-optical NAFs via TO nonlinear effect. A microdisk resonator accumulates optical energy at its resonant wavelength. With high input power at resonant wavelength, the power density inside the microdisk resonator will be amplified substantially because of its resonant enhancement property and ultra-small geometric dimensions, which therefore induces TO effect in silicon waveguide and causes a red resonance shift. The optical nonlinear effects include the Kerr effect, TPA, FCA, free-carrier dispersion (FCD), and carrier-related TO effect in the silicon microdisk. With continuous light input, the TO effect significantly dominates over other nonlinear effects. We measure the transmission spectra and nonlinear transfer functions at the through and drop ports of the fabricated microdisk modulator. As can be seen in Figs. 7(a) and 7(b), the resonant peak redshifts with the increase of input optical power. We keep the input light wavelength fixed to the initial resonant wavelength, and measure the output power of the through port with the change of input power. As shown in Fig. 7(c), the output optical power of the through port first increases linearly with the input optical power, and then gradually rapidly. In Fig. 7(c), the power input and output are plotted in milliwatts, and the inset shows the trend for the input optical power less than 0.1 mW. The nonlinear threshold is measured to be 0.082 mW. We further demonstrate radial basis function by tuning the input light wavelength slightly above the initial resonant peak. As shown in Fig. 7(d), different nonlinear curves are measured by adjusting the wavelength detuning. It is noteworthy that the bistability only occurs at the wavelength far away from the resonant wavelength. When the wavelength detuning is small, the bistability has no impact on the reliability of the measured nonlinear functions. As shown in Figs. 7(e) and 7(f), different NAFs are realized at the drop port by tuning the wavelength. It is noteworthy that although our measured NAFs are not the exact normally used NAFs, they can be regarded as an approximation. And these measured NAFs also perform well in neural networks.

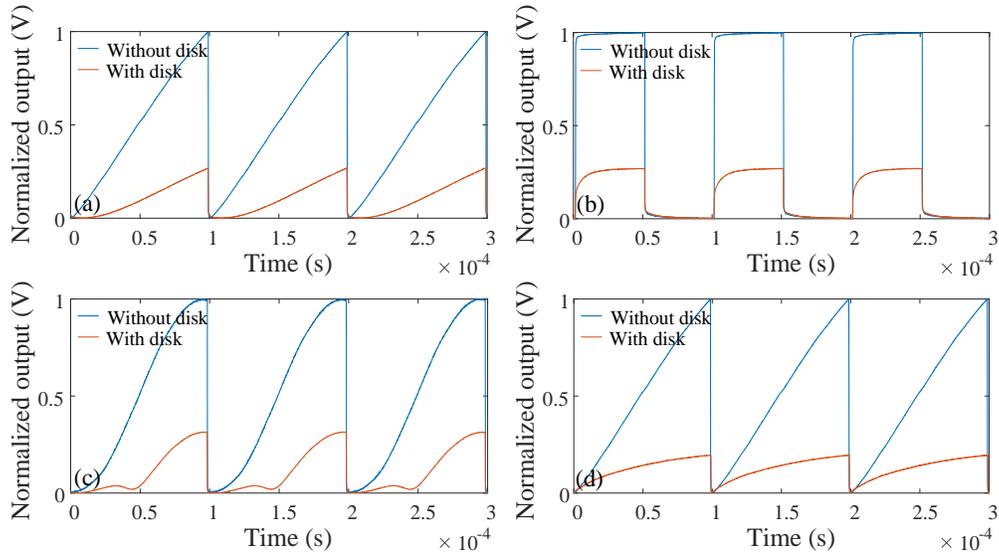

Fig. 8. (a)(b) Measured waveforms at the through port after MZM sawtooth signal and square wave signal modulation. (c) Measured waveform of the through port with wavelength detuning of 30 pm. (d) Measured waveform of the drop port after MZM sawtooth modulation.

We then measure the response speed of the implemented all-optical NAFs. The input light beam is first modulated by an intensity modulator loaded with a sawtooth or square electrical signal before entering the device. The output light from the device is detected by a PD and then captured by an oscilloscope. As shown in Fig. 8(a), we send a sawtooth signal into the device and measure the output signal of the through port, which agrees well with the measured nonlinear response in Fig. 7(c). Furthermore, to measure the accurate response time, we send a square wave signal into the device and get the output signal shown in Fig. 8(b). We get the rise and fall time of ~12 μs, which is the time that output signal takes to become stable when the input signal changes its level. Figure 8(c) shows the nonlinear response at the through port with a wavelength detuning of 30 pm, which is in good agreement with the result in Fig. 7(d). In Fig. 8(d), the measured nonlinear response of the drop port also agrees well with the result in Fig. 7(e) when sending a sawtooth signal into the device.

To verify the effectiveness of the implemented NAFs, we perform a simulation of convolutional neural network (CNN) based-MNIST handwritten digit classification. As shown in Fig. 9(a), the input image of a handwritten digit is a 2D matrix comprising 28×28 pixels. The experimentally measured all-optical nonlinear activation function (clamped ReLU) in Fig. 7(e) is used. The input undergoes convolutions, experimental ReLU5 operations, and pooling (maxpool), followed by two fully connected layers. We also adopt Linear rectification function (ReLU) as a comparison of optical activation function. The MNIST dataset is divided into training and test samples, with batch sizes of 32 and 2000, respectively. The simulated model is built in PyTorch, and it is trained by the AdamOptimizer with a default learning rate of 0.001. Figure 9(b) shows the negative log-likelihood loss as a function of samples fed into the network, during the training and test stages. The network is found to converge within 1 epoch to negligible loss and 98% accuracy during testing for both NAFs. As can be seen, this network converges faster when using the all-optical nonlinear activation function. Figure 9(c) shows a sample of handwritten digit inputs, with labels correctly predicted by the network.

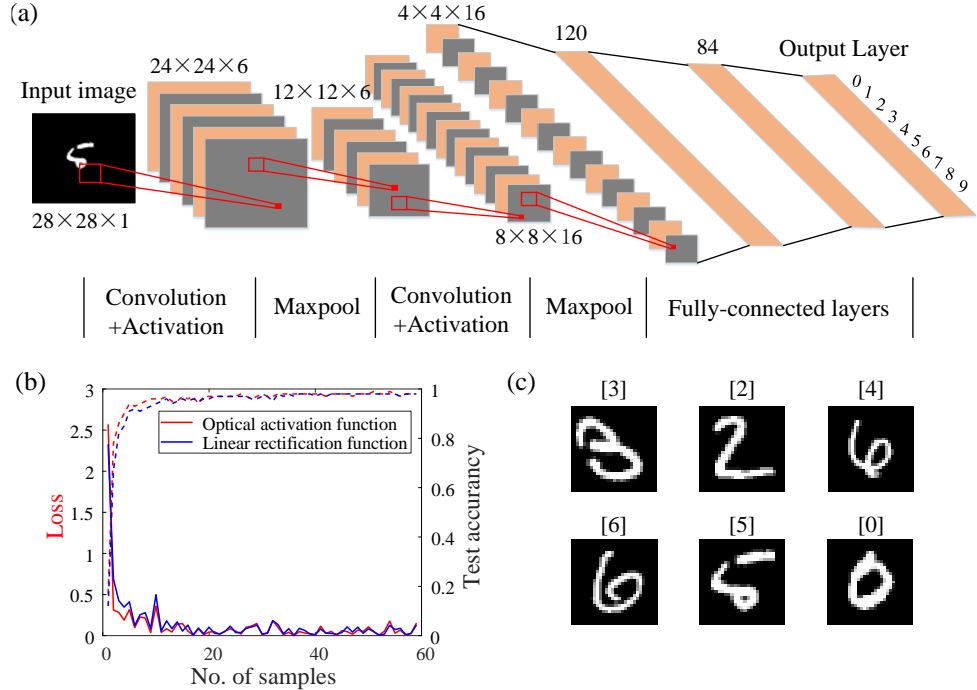

Fig. 9. (a) Schematic of MNIST handwritten digit classification CNN, comprising convolution, maxpool, experimental nonlinear function, and two fully connected layers, resulting in outputs in the range of [0, 9]. (b) Negative log likelihood loss (left axis) during training and testing stages, and accuracy during testing (right axis) as a function of number of samples. (c) Example of network predicted labels corresponding to six input images.

## 5. Discussion

Table1 shows the detailed comparison of different approaches based on integrated platforms. The add-drop microdisk modulator has a low all-optical nonlinear threshold, which provides an opportunity for implementation multiple E-O and all-optical NAFs. In the experiment, we demonstrate three typical E-O NAFs, and other three types nonlinear responses could be also implemented by further tuning the device [23]. The dynamic evolution of the nonlinear responses is also illustrated in the experiment. The proposed add-drop silicon microdisk modulator has a nonlinear E-O response speed of 1 GHz due to free-carrier injection modulation. Four different all-optical NAFs are also realized. The all-optical nonlinear response speed is less than 100 kHz due to the TO effect, which could be further improved to realize a much higher response speed up to GHz by using FCD or Kerr effect [29, 30].

With further improvements, a microheater could be integrated on the microdisk modulator to precisely control the resonance peak. Thus, both E-O and all-optical NAFs could be programmable and reconfigurable [29, 30]. In addition, it should be noted that the optical bistability curve strongly depends on the wavelength detuning between the input optical source and the initial resonant peak, which may be not conducive to the realization of stable NAFs [41, 42]. To solve this problem, we implement stable all-optical NAFs by controlling the wavelength detuning. In the experiment, the input light wavelength is slightly off or aligned with the initial resonant peak of the add-drop microdisk resonator, which could greatly reduce the effect of bistability and provide a stable nonlinear response. Furthermore, self-pulsation could be also suppressed in the silicon MDR modulator based on free-carrier depletion [50].

Table1. Detailed comparison of different approaches based on photonic integrated platforms

| Technologies | Materials | Types | Speed | Mechanism |
|---|---|---|---|---|
| PD-Modulator [23] | Si | 6 | GHz | E-O |
| MZI [24] | SiN | 2 | GHz | E-O |
| EA modulator [25] | ITO-Si | 1 | GHz | E-O (FCA) |
| MZI-MRR [29] | Si | 4 | GHz | O (FCD effect) |
| MZI-MRR [30] | SiN | 4 | 10 GHz | O (Kerr effect) |
| Ge/Si waveguide [31] | Ge/Si | 1 | 70 MHz | O (FCD) |
| Ge/Si microring [32] | Ge/Si | 3 | <100 kHz | O (TO effect) |
| MZI mesh [35] | Si | 2 | / | O (Interference) |
| Si microring [36] | Si | 4 | <100 kHz | O (TO effect) |
| This work | Si | 6 | GHz | E-O |
| This work | Si | 4 | <100 kHz | O (TO effect) |

"O" refers to all-optical implementation of nonlinear activation function.

## 6. Conclusion

In summary, we propose and demonstrate an ultra-compact integrated microdisk modulator to achieve multiple NAFs based on silicon photonic platform. Based on the plasma dispersion effect, the fabricated add-drop microdisk modulator exhibits a variety of high-speed nonlinear E-O transfer functions including sigmoid, radial basis, and ReLU by electrically controlling free-carrier injection modulation. The measured E-O nonlinear response speed is up to 1 GHz. Meanwhile, four different all-optical nonlinear responses are also realized by employing the TO nonlinear effect of the silicon microdisk resonator. The measured all-optical nonlinear response speed is about 100 kHz. Moreover, we simulate a benchmark task, a CNN based-handwritten digit classification with an experimentally measured activation function and obtain an accuracy of 98%. The demonstrated microdisk modulator holds great potential for the implementation of large-scale PNNS.

**Declaration of competing interest**

The authors declare that they have no known competing financial interests or personal relationships that could have appeared to influence the work reported in this paper.

**Acknowledgements**

This work was supported by the National Key R&D Program of China (No. 2018YFE0201800), and the National Natural Science Foundation of China (NSFC) (62105028, 62071042, 62005018).

## References

1. Y. LeCun, Y. Bengio, and G. Hinton, "Deep learning," Nature **521**, 436-444 (2015).
2. S. Grigorescu, B. Trasnea, T. Cocias and G. Macesanu, "A survey of deep learning techniques for autonomous driving," J. Field Robot. **37**, 362–386 (2020).
3. H. Chen, Y. Zhang, M. K. Kalra, F. Lin, Y. Chen, P. Liao, J. Zhou and G. Wang, "Low-dose CT with a residual encoder-decoder convolutional neural network," IEEE Trans. Med. Imaging **36**, 2524–2535 (2017).
4. I. A. Basheer, and M. Hajmeer, "Artificial neural networks: fundamentals, computing, design, and application," J. Microbiol Methods **43**(1), 3-31 (2000).
5. R. Preissl, T. M. Wong, P. Datta, M. Flickner, R. Singh, S. K. Esser, W. P. Risk, H. D. Simon, and D. S. Modha, "Compass: a scalable simulator for an architecture for cognitive computing," in SC '12: Proc. Int. Conf. on High Performance Computing, Networking, Storage and Analysis (IEEE, 2012), pp. 1-11.
6. J. von Neumann, The Computer and the Brain (Yale Univ. Press, 1958).
7. C. Li, X. Zhang, J. Li, T. Fang, and X. Dong, "The challenges of modern computing and new opportunities for optics," PhotoniX **2**(1), 1-31 (2021).
8. Y. Shen, N. C. Harris, S. Skirlo, M. Prabhu, T. Baehr-Jones, M. Hochberg, and M. Soljačić, "Deep learning with coherent nanophotonic circuits," Nat. Photonics **11**(7), 441-446 (2017).
9. J. Feldmann, N. Youngblood, C. D. Wright, H. Bhaskaran and W. H. Pernice, "All-optical spiking neurosynaptic networks with self-learning capabilities," Nature **569**(7755), 208-214 (2019).
10. J. Feldmann, N. Youngblood, M. Karpov, H. Gehring, X. Li, M. Stappers, and H. Bhaskaran, "Parallel convolutional processing using an integrated photonic tensor core," Nature **589**(7840), 52-58 (2021).
11. X. Xu, M. Tan, B. Corcoran, J. Wu, A. Boes, T. G. Nguyen, and D. J. Moss, "11 TOPS photonic convolutional accelerator for optical neural networks," Nature **589**(7840), 44-51 (2021).
12. H. Zhang, M. Gu, X. D. Jiang, J. Thompson, H. Cai, S. Paesani, and A. Q. Liu, "An optical neural chip for implementing complex-valued neural network," Nat. Commun. **12**(1), 1-11 (2021).
13. C. Wu, H. Yu, S. Lee, R. Peng, I. Takeuchi, and M. Li, "Programmable phase-change metasurfaces on waveguides for multimode photonic convolutional neural network," Nat. Commun. **12**(1), 1-8 (2021).
14. S. Xu, J. Wang, H. Shu, Z. Zhang, S. Yi, B. Bai, and W. Zou, "Optical coherent dot-product chip for sophisticated deep learning regression," Light: Sci. Appl. **10**(1), 1-12 (2021).
15. C. Huang, S. Fujisawa, T. F. de Lima, A. N. Tait, E. C. Blow, Y. Tian, and P. R. Prucnal, "A silicon photonic–electronic neural network for fibre nonlinearity compensation," Nat. Electron. **4**(11), 837-844 (2021).
16. C. Huang, V. J. Sorger, M. Miscuglio, M. Al-Qadasi, A. Mukherjee, L. Lampe, and B. J. Shastri. "Prospects and applications of photonic neural networks," ADV PHYS-X **7**(1), 1981155 (2022).
17. B. J. Shastri, A. N. Tait, T. Ferreira de Lima, W. H. Pernice, H. Bhaskaran, C. D. Wright, and P. R. Prucnal, "Photonics for artificial intelligence and neuromorphic computing," Nat. Photonics **15**(2), 102-114 (2021).
18. H. Zhou, J. Dong, J. Cheng, W. Dong, C. Huang, Y. Shen, and X. Zhang, "Photonic matrix multiplication lights up photonic accelerator and beyond," Light: Sci. Appl. **11**(1), 1-21 (2022).
19. M. Mario, M. Armin, H. Zibo, I. A. Shaimaa, G. Jonathan, V. K. Alexander, P. Matthew, and J. S. Volker, "All-optical nonlinear activation function for photonic neural networks," Opt. Mater. Express **8**, 3851-3863 (2018).
20. Y. Zuo, B. Li, Y. Zhao, Y. Jiang, Y. C. Chen, P. Chen, and S. Du, "All-optical neural network with nonlinear activation functions," Optica **6**(9), 1132-1137 (2019).
21. G. Mourgias-Alexandris, A. Tsakyridis, N. Passalis, A. Tefas, K. Vyrsokinos, and N. Pleros, "An all-optical neuron with sigmoid activation function," Opt. Express **27**, 9620-9630 (2019).
22. I. A. Williamson, T. W. Hughes, M. Minkov, B. Barlett, S. Pai, and S. Fan, "Reprogrammable electro-optic nonlinear activation functions for optical neural networks," IEEE J. Quantum Electron. **26**(1), 1-12 (2019).
23. A. N. Tait, T. F. De Lima, M. A. Nahmias, H. B. Miller, H. T. Peng, B. J. Shastri, and P. R. Prucnal, "Silicon photonic modulator neuron," Phys. Rev. Appl. **11**(6), 064043 (2019).
24. M. M. P. Fard, I. A. D. Williamson, M. Edwards, K. Liu, S. Pai, B. Bartlett, M. Minkov, T. W. Hughes, S. Fan, and T. Nguyen, "Experimental realization of arbitrary activation functions for optical neural networks," Opt. Express **28**, 12138-12148 (2020).
25. R. Amin, J. K. George, S. Sun, T. Ferreira de Lima, A. N. Tait, J. B. Khurgin, and V. J. Sorger, "ITO-based electro-absorption modulator for photonic neural activation function", APL Mater. **7**(8), 081112 (2019).
26. J. K. George, A. Mehrabian, R. Amin, J. Meng, T. F. Lima, A. N. Tait, B. J. Shastri, T. El-Ghazawi, P. R. Prucnal, and V. J. Sorger, "Neuromorphic photonics with electro-absorption modulators," Opt. express **27**(4): 5181-5191 (2019).
27. T. S. Rasmussn, Y. Yu, and J. Mork, "All-optical non-linear activation function for neuromorphic photonic computing using semiconductor Fano lasers," Opt. Lett. **45**(14), 3844-3847 (2020).
28. R. Amin, J. K. George, S. Sun, T. Ferreira de Lima, A. N. Tait, J. B. Khurgin, and V. J. Sorger, "ITO-based electro-absorption modulator for photonic neural activation function," APL Materials, **7**(8), 081112 (2019).
29. A. Jha, C. Huang, and P. R. Prucnal, "Reconfigurable all-optical nonlinear activation functions for neuromorphic photonics," Opt. Lett. **45**(17), 4819-4822 (2020).
30. A. Jha, C. Huang, and P. R. Prucnal, "Programmable, high-speed all-optical nonlinear activation functions for neuromorphic photonics," in Optical Fiber Communication Conference (OSA, 2021), pp. Tu5H-3.
31. H. Li, B. Wu, W. Tong, J. Dong, and X. Zhang, "All-Optical Nonlinear Activation Function Based on Germanium Silicon Hybrid Asymmetric Coupler," IEEE J. Sel. Top. Quantum Electron. **29**(2), 1-6 (2023).

32. B. Wu, H. Li, W. Tong, J. Dong, and X. Zhang, "Low-threshold all-optical nonlinear activation function based on a Ge/Si hybrid structure in a microring resonator," Opt. Mater. Express **12**(3), 970-980 (2022).
33. C. R. Huang, T. Ferreira de Lima, A. Jha, S. Abbaslou, A. N. Tait, B. J. Shastri, and P. R. Prucnal, "Programmable silicon photonic optical thresholder," IEEE Photonics Technol. Lett. **31**, 1834–1837 (2019).
34. C. R. Huang, A. Jha, T. Ferreira de Lima, A. N. Tait, B. J. Shastri, and P. R. Prucnal, "On-chip programmable nonlinear optical signal processor and its applications," IEEE J. Sel. Top. Quantum Electron. **27**, 6100211 (2021).
35. Q. Li, S. Liu, Y. Zhao, W. Wang, Y. Tian, J. Feng, and J. Guo, "Optical Nonlinear Activation Functions Based on MZI-Structure for Optical Neural Networks," in 2020 Asia Communications and Photonics Conference (ACP) and International Conference on Information Photonics and Optical Communications (IPOC) (IEEE, 2020), pp. 1–3.
36. W. Yu, S. Zheng, Z. Zhao, B. Wang, and W. Zhang, "Reconfigurable low-threshold all-optical nonlinear activation functions based on an add-drop silicon microring resonator," IEEE Photon. J. **14**(6), 1-7 (2022).
37. X. Qian, M. Sasikanth, S. Brad, S. Jagat, and L. Michal, "12.5 Gbit/s carrier-injection based silicon micro-ring silicon modulators," Opt. Express **15**, 430 (2007).
38. W. Zhang, and J. Yao, "Electrically tunable silicon-based on-chip microdisk resonator for integrated microwave photonic applications," APL Photonics **1**(8), 080801 (2016).
39. J. Sun, R. Kumar, M. Sakib, J. B. Driscoll, H. Jayatilleka, and H. Rong, "A 128 Gb/s PAM4 silicon microring modulator with integrated thermo-optic resonance tuning," IEEE J. Light. Techol. **37**(1), 110-115 (2018).
40. Y. Zhang, H. Zhang, M. Li, P. Feng, L. Wang, X. Xiao, and S. Yu, "200 Gbit/s optical PAM4 modulation based on silicon microring modulator," in 2020 European Conference on Optical Communications (ECOC) (IEEE, 2020), pp. 1-4.
41. V. R. Almeida, and M. Lipson, "Optical bistability on a silicon chip," Opt. Lett. **29**(20), 2387-2389 (2004).
42. Q. Xu, and M. Lipson, "Carrier-induced optical bistability in silicon ring resonators," Opt. Lett. **31**(3), 341-343 (2006).
43. J. J. Hopfield and D. W. Tank, "Neural computation of decisions in optimization problems," Biol. Cybern. **52**(3), 141 (1985).
44. C. Cortes, and V. Vapnik, "Support-vector networks," Mach. Learn. **20**(3), 273-297 (1995).
45. U. P. Wen, K. M. Lan, and H. S. Shih, "A review of Hopfield neural networks for solving mathematical programming problems," European Journal of Operational Research **198**(3), 675-687 (2009).
46. Y. Huang, W. Wang, L. Qiao, X. Hu, and T. Chu, "Programmable low-threshold optical nonlinear activation functions for photonic neural networks," Opt. Lett. **47**(7), 1810-1813 (2022).
47. X. Hu, D. Wu, H. Zhang, W. Li, D. Chen, L. Wang, X. Xiao, and S. Yu, "High-speed and high-power germanium photodetector with a lateral silicon nitride waveguide," Photon. Res. **9**, 749-756 (2021).
48. X. Hu, D. Wu, D. Chen, L. Wang, X. Xiao, and S. Yu, "180 Gbit/s $Si_3N_4$-waveguide coupled germanium photodetector with improved quantum efficiency," Opt. Lett. **46**, 6019-6022 (2021).
49. X. Hu, H. Zhang, D. Wu, D. Chen, L. Wang, X. Xiao, and S. Yu, "High-performance germanium avalanche photodetector for 100 Gbit/s photonics receivers," Opt. Lett. **46**, 3837-3840 (2021).
50. W. Shen, G. Zhou, J. Du, L. Zhou, K. Xu, and Z. He, "High-speed silicon microring modulator at the 2 μm waveband with analysis and observation of optical bistability," Photon. Res. **10**, A35-A42 (2022).